\documentclass[
    ,draft   
    ,4apaper          
    ,numberedheadings
  ]
  {aipproc}

\layoutstyle{6x9}
\usepackage[english]{babel} 
\usepackage{graphicx}
\usepackage{bbm,amssymb,amsopn}
\usepackage{dsfont}

\begin{document}

\bibliographystyle{aipproc}

\title{Heat transport in quantum harmonic chains with Redfield baths}
 
\author{Bojan \v{Z}unkovi\v{c}}{
address={Department of Physics, Faculty of Mathematics and Physics, University of Ljubljana, Jadranska 19, SI-1000 Ljubljana, Slovenia},
email={bojan.zunkovic@fmf.uni-lj.si}}

\author{Toma\v z Prosen}{
address={Department of Physics, Faculty of Mathematics and Physics, University of Ljubljana, Jadranska 19, SI-1000 Ljubljana, Slovenia},
email={tomaz.prosen@fmf.uni-lj.si}}

\date{\today}

\begin{abstract}
We provide an explicit method for solving general markovian master equations for quadratic bosonic Hamiltonians with linear bath operators. As an example we consider a one-dimensional quantum harmonic oscillator chain coupled to thermal reservoirs at both ends of the chain. We derive an analytic solution of the Redfield master equation for homogeneous harmonic chain and recover classical results, namely, vanishing temperature gradient and constant heat current in the thermodynamic limit. In the case of the disordered gapped chains we observe universal heat current scaling independent of the bath spectral function, the system-bath coupling strength, and the boundary conditions.
\end{abstract}



\maketitle
\section{Introduction}
The question of the microscopic origin of well established Fourier's law of heat conduction has remained unsolved for many decades \cite{giulio,ford,bonetto,lepri,david,dhar08}. The main issue is to find the necessary and sufficient conditions for normal transport. Therefore, it is important to understand basic, simple models, that are expected to embrace important features of complicated, more realistic models.  It is believed that for solid systems disorder and anharmonicity play a key role in determining the transport properties. Focusing on the quantum of quantum mechanical systems, this encouraged the development of different formalisms for treating large non-equilibrium quantum systems, such as the Keldysh technique \cite {keldysh}, Landauer-Butikker formalism \cite{landauer}, quantum Langevin equation \cite{qle}, quantization in the Fock space of operators \cite{njp,njp2,pro10,prosel10} giving insight into the region far from equilibrium. 

The simplest, but nevertheless non-trivial model in the context of heat transport properties, is the coupled one-dimensional harmonic oscillator chain. Many different methods on modeling the heat reservoirs have been presented. Much effort has been given to solve the classical many-body Langevin equation, or the corresponding master (Liouville) equation. For the homogeneous harmonic oscillator chain size independent heat flux and vanishing temperature gradient have been proven \cite{lieb}. On the other hand, a model dependent scaling of the heat flux has been found for the disordered chain \cite{dharprl}. Dhar and Shastry have derived a generalized quantum Langevin equation by using the Ford-Kac-Mazur formalism \cite{dhar}, and have got similar results as in the classical case. Further, it has been argued in detail that the heat and particle transport properties of the model depend on the spectral properties of the heat bath \cite{dhar08} and boundary conditions of the model, in the case of a gapless harmonic oscillator chain, i.e. with vanishing on-site potential. In Ref. \cite{gaubut07}  ordered and disordered one-dimensional harmonic chains within the quantum mechanical Langevin equation have been analyzed. No clear statement on scaling of the heat current in the disordered chain has been given. The connection to the entanglement has been observed as well.

In this paper we consider a different, effective approach of open quantum systems \cite{breuer} in the many-body context, which in the past lead us to interesting and fruitful results in the realm of one-dimensional fermionic models \cite{njp,njp2} and can be used to investigate heat transport properties via the Redfield master  equation \cite{saito,satami00}. It has been shown \cite{prosel10} that the Lindblad master equation, a special case of the Redfield master equation, can be solved exactly for quadratic bosonic models. We shall see that similar formalism can be used to get the non-equilibrium stationary state (NESS) of the Redfield master equation for a general quadratic boson Hamiltonian with linear bath operators. We further apply the method to one-dimensional harmonic oscillator chain, which serves as a well known model and can be used to test our approach and possibly give an alternative description of heat baths.

This paper is organized as follows. In section \ref{sec2.0} we review the main concepts of the quantization in the Fock space of operators for bosonic systems \cite{prosel10}. We show that this method can be used to solve the Redfield master equation and derive the Lyapunov equation for the covariance matrix, which determines the NESS. In section \ref{sec3.0} we consider the one-dimensional harmonic oscillator chain model in the context of the Redfield master equation. First, in subsection 3.1 we describe a general method of solution of the open harmonic chain and in subsections 3.2, 3.3 provide two explicit solutions for the homogeneous chain, namely a compact solution for the case of symmetric (left/right) system-bath coupling and ohmic bath correlation function and a recursive perturbative (in the coupling) solution, again for symmetric coupling strengths and a general bath correlation function. In both cases simple expressions for the heat conductivity are obtained. Further, in 
subsection 3.4 we introduce disorder and find power law scaling of heat current, that is independent of the fine reservoir properties, i.e. the spectral function of the bath, and the coupling strength.
In section 4 we summarize and conclude.

This paper mainly presents original research results which have not been published before, however we review some material from \cite{prosel10} and \cite{njp,njp2} in order to make the manuscript more smoothly readable and self-contained. 

\section{Exact solution of the Redfield equation for quadratic boson system}
\label{sec2.0}
Our aim in this section is to extend operator Fock space quantization for bosonic open quantum system. We focus on the Redfield master equation since the Lindblad case can be considered as a limit of equal-time bath correlation functions similar as in Ref. \cite{njp2}. We shall provide an exact solution of the Liouville master equation in the Redfield form
\begin{eqnarray}
\label{eq:master}
&\frac{{\rm d}}{{\rm d}t}\rho(t)=\hat{\mathcal{L}}\rho(t),\\ \label{eq:Liouville}
&\hat{\mathcal{L}}\rho={\rm i}[\rho,H]+\hat{\mathcal{D}}\rho,\\ \label{eq:dissipator} 
&\hat{\mathcal{D}}\rho=\sum_{\mu\nu}\int_0^\infty\mathrm{d}t\Gamma_{\nu,\mu}(\tau)[X_\mu(-\tau)\rho,X_\nu]+{\rm h.c.}
\end{eqnarray}
The Liouvillean (\ref{eq:Liouville}) includes two parts. The first part is the generator of the unitary evolution and is determined by the Hamiltonian $H$. The second part, the dissipator, introduces effectively the influence of infinite large environments (reservoirs or baths) with the correlation functions $\Gamma_{\nu,\mu}(t)$. 
We assume a general quadratic Hamiltonian and linear coupling operators
\begin{eqnarray}
\label{eq:H_X_def}
&H=\underline{p}\cdot{\bf P}\underline{p}+\underline{q}\cdot{\bf Q}\underline{q}+\underline{p}\cdot{\bf R}\underline{q}+\underline{q}\cdot{\bf R}\underline{p},\\ \nonumber
&X_{\nu}=\underline{x}_{\nu}^{\rm q}\cdot\underline{q}+\underline{x}_{\nu}^{\rm p}\cdot\underline{p},
\end{eqnarray}
where  ${\bf P, Q, R}$ are real, symmetric matrices, $p_j, q_j$ are momentum and coordinate operators, respectively, and $x^{\rm p}_{\nu,j}, x^{\rm q}_{\nu,j}$ are real numbers (coupling amplitudes), which determine the coupling operators $X_\nu$. Greek subscript letters are used to denote different reservoirs and latin subscript letters denote the position in the system (site index $j=1,2,\ldots n$) of size $n$. We shall use this notation throughout the paper, as well as bold latin letters for matrices (${\bf A}$) and the hat for super-operators ($\hat{a}$). In the equation (\ref{eq:H_X_def}) we introduced the following notation of a column vector $\underline{x}=(x_1, x_2,\ldots x_n)^{\rm T}$, where $x$ can represent a number, an operator or a super-operator, which is evident from the context, as well as a dot product that assigns only transposition and {\it not} the complex conjugate of the left operand $\underline{x}\cdot\underline{y}=\sum_{j=1}^n x_jy_j$.
 
The problem could equally well be formulated in the second quantization picture using the creation and annihilation operators $a^\dag_j, a_j$. It is obvious that the two representations are equivalent, however, using the Hermitian operators results in a real Lyapunov equation for the stationary correlation matrix. This simplifies the algebra and the numerical calculations.
\subsection{Coordinate-momentum picture super-operators}
We conform to Ref. \cite{prosel10},  where the solution of the Lindblad master equation has been found exploiting the algebra in a pair of dual vector spaces $\mathcal{K}$ and $\mathcal{K}'$, where $\mathcal{K}$ contains trace class operators (density matrices) and $\mathcal{K}'$ contains unbounded operators (physical observables) \footnote{For details see Ref. \cite{prosel10}.}. We will adopt the Dirac notation and write an element of $\mathcal{K}'$ as {\it ket} $|\rho\rangle$ and an element of $\mathcal{K}$ as {\it bra} $(A|$, so that their contractions give the expectation value of $A$ for the state $\rho$
\begin{equation}
(A|\rho\rangle={\rm tr}\{ A\rho\}.
\label{eq:inner_prod}
\end{equation}
The basic tool used in this section are left and right multiplication maps over $\mathcal{K}$ defined as
\begin{equation}
\label{eq:mult_maps}
\hat{b}^{\rm L}|\rho\rangle=|b\rho\rangle, \quad \hat{b}^{\rm R}=|\rho b\rangle.
\end{equation}
From the cyclicity of the trace and the definition of the inner product (\ref{eq:inner_prod}) we deduce the action of the maps (\ref{eq:mult_maps}) on the elements of the adjoint space
\begin{equation}
(A|\hat{b}^{\rm L}=(Ab|,\quad (A|\hat{b}^{\rm R}=(bA|.
\end{equation}
Further we define $4n$ maps $\hat{c}_{\nu,j},\hat{c}_{\nu,j}'$ for $j=1,2,\ldots,n$ and $\nu=0,1$
\begin{eqnarray}
\label{eq:def_c}
&\hat{c}_{0,j}=\hat{p}_j^{\rm L}, \quad \hat{c}_{0,j}'={\rm i}(\hat{q}_j^{\rm L}-\hat{q}_j^{\rm R}),\\ \nonumber
&\hat{c}_{1,j}=\hat{q}_j^{\rm R}, \quad \hat{c}_{1,j}'={\rm i}(\hat{p}_j^{\rm R}-\hat{p}_j^{\rm L})
\end{eqnarray}
satisfying almost canonical commutation relations
\begin{eqnarray}
[\hat{c}_{\mu,j},\hat{c}_{\nu,k}]=[\hat{c}_{\mu,j}',\hat{c}_{\nu,k}']=0,\quad [\hat{c}_{\mu,j},\hat{c}_{\nu,k}']=\delta_{j,k}\delta_{\mu,\nu}.
\end{eqnarray}
Note that $\hat{c}_{\mu,j}'\neq\hat{c}_{\mu,j}^\dag$. Another important property is that the maps $\hat{c}'_{\nu,j}$ left-annihilate the identity operator $(1|\hat{c}'_{\nu,j}=1$. We shall now formulate the Liouvillean as a quadratic form in the maps $\hat{c}_{\nu,j}, \hat{c}_{\nu,j}'$ using the inverse relations
\begin{eqnarray}
\label{eq:inv_rel3}
\hat{p}_j^{\rm L}=\hat{c}_{0,j},& \quad \hat{q}_j^{\rm L}=\hat{c}_{1,j}-{\rm i}\hat{c}_{0,j}',\\ \nonumber
\hat{q}_j^{\rm R}=\hat{c}_{1,j},& \quad \hat{p}_j^{\rm R}=\hat{c}_{0,j}-{\rm i}\hat{c}_{1,j}'.
\end{eqnarray}
Let us first rewrite the contribution of the Hamiltonian to the Liouvillean
\begin{eqnarray}
\label{eq:liouv_unit}
&{\rm i}(\hat{H}^{\rm R}-\hat{H}^{\rm L})={\rm i}(\underline{\hat{p}}^{\rm R}\cdot{\bf P}\underline{\hat{p}}^{\rm R}+\underline{\hat{q}}^{\rm R}\cdot{\bf Q}\underline{\hat{q}}^{\rm R}+\underline{\hat{q}}^{\rm R}\cdot{\bf R}\underline{\hat{p}}^{\rm R}+\underline{\hat{p}}^{\rm R}\cdot{\bf R}\underline{\hat{q}}^{\rm R}\\ \nonumber
&~-\underline{\hat{p}}^{\rm L}\cdot{\bf P}\underline{\hat{p}}^{\rm L}-\underline{\hat{q}}^{\rm L}\cdot{\bf Q}\underline{\hat{q}}^{\rm L}-\underline{\hat{p}}^{\rm L}\cdot{\bf R}\underline{\hat{q}}^{\rm L}-\underline{\hat{q}}^{\rm L}\cdot{\bf R}\underline{\hat{p}}^{\rm L})\\ \nonumber
&=2(\underline{\hat{c}}_{1}'\cdot{\bf P}\underline{\hat{c}}_{0}-\underline{\hat{c}}_{0}'\cdot{\bf Q}\underline{\hat{c}}_{1}+\underline{\hat{c}}_{1}'\cdot{\bf R}\underline{\hat{c}}_{1}-\underline{\hat{c}}_{0}'\cdot{\bf R}\underline{\hat{c}}_{0})+{\rm i}(-\underline{\hat{c}}_{1}'\cdot{\bf P}\underline{\hat{c}}_{1}'+\underline{\hat{c}}_{0}'\cdot{\bf Q}\underline{\hat{c}}_{0}').
\end{eqnarray}
Above, anti-Hermitian part of the Liouvillean (\ref{eq:liouv_unit}) left-annihilates the identity operator and is quadratic in the maps $\hat{c}_{\nu,j}, \hat{c}_{\nu,j}'$. We shall shortly see that these characteristics are satisfied for the second, dissipative part of the Liouvillean (\ref{eq:dissipator}) as well. The dissipator can be represented in the following, useful form
\begin{equation}
\label{eq:dissip_def}
\hat{\mathcal{D}}=\sum_{\mu,\nu}\sum_{l,m}^{\{{\rm p,q}\}}\sum_{j,k=1}^nx^m_{\nu,k}\int_0^\infty{\rm d}\tau f^l_{\mu,j}(-\tau)\left( \Gamma^\beta_{\nu,\mu}(\tau)\hat{\mathcal{L}}^{l,m}_{j,k}+\Gamma^{\beta*}_{\nu,\mu}(\tau)\hat{\mathcal{R}}^{l,m}_{j,k}\right).
\end{equation}
We introduced the Heisenberg propagator
\begin{equation}
\label{eq:heiss_prop_def}
\underline{x}_\nu=(\underline{x}_\nu^{\rm p},\underline{x}_\nu^{\rm q}),\quad
\underline{f}_\nu^{\rm a}(t)=\underline{x}_\nu^{\rm a}\cdot\exp({-{\rm i~ad}\,H}t),\quad {\rm a\in\{p,q\},}
\end{equation}
and the fundamental dissipators
\begin{eqnarray}
\hat{\mathcal{L}}^{\rm a,b}_{j,k}|\rho\rangle=|[a_j,\rho b_k]\rangle,\quad \hat{\mathcal{R}}^{\rm a,b}_{j,k}|\rho\rangle=|[b_k\rho, a_j]\rangle, 
\quad j,k=1,\ldots n,
\end{eqnarray}
where $a_j = p_j$ if $\rm a = p$ or $a_j = q_j$ if $\rm a = q$ and 
$b_j = p_j$ if $\rm b = p$ or $b_j = q_j$ if $\rm b = q$. Expressing the relations (\ref{eq:inv_rel3}) we obtain after some algebra a compact form of the fundamental dissipators 
\begin{eqnarray}
\label{eq:fund_dissip}
&\hat{\mathcal{L}}^{\rm q,q}_{j,k}={\rm i}\hat{c}_{0,k}'(\hat{c}_{1,j}-{\rm i}\hat{c}_{0,j}'),\qquad\hat{\mathcal{R}}^{\rm q,q}_{j,k}=-{\rm i}\hat{c}_{0,k}'\hat{c}_{1,j},\\ \nonumber
&\hat{\mathcal{L}}^{\rm p,p}_{j,k}=-{\rm i}\hat{c}_{1,k}'\hat{c}_{0,j},\qquad\hat{\mathcal{R}}^{\rm p,p}_{j,k}={\rm i}\hat{c}_{1,k}'(\hat{c}_{0,j}-{\rm i}\hat{c}_{1,j}'),\\ \nonumber
&\hat{\mathcal{L}}^{\rm p,q}_{j,k}={\rm i}\hat{c}_{0,k}'\hat{c}_{0,j},\qquad\hat{\mathcal{R}}^{\rm p,q}_{j,k}=-{\rm i}\hat{c}_{0,k}'(\hat{c}_{0,j}-{\rm i}\hat{c}_{1,j}'),\\ \nonumber
&\hat{\mathcal{L}}^{\rm q,p}_{j,k}=-{\rm i}\hat{c}_{1,k}'(\hat{c}_{1,j}-{\rm i}\hat{c}_{0,j}'),\qquad\hat{\mathcal{R}}^{\rm q,p}_{j,k}={\rm i}\hat{c}_{1,k}'\hat{c}_{1,j}.
\end{eqnarray}
Let us now rearrange and group the terms in the dissipator (\ref{eq:dissip_def}) according to their meaning.  All information about the baths can be encoded in the bath matrices
\begin{eqnarray}
\mathbf{M}^{m,l}=\sum_{\mu,\nu}\int_0^\infty{\rm d}\tau \left(\underline{x}^m_{\nu}\otimes\underline{f}_{\mu}^l(-\tau)\right)\Gamma^\beta_{\nu,\mu}(\tau), \quad l,m\in\{{\rm p,q}\},
\label{eq:bath_matrix_def}
\end{eqnarray}
which simplifies the equation (\ref{eq:dissip_def}) for the dissipator
\begin{eqnarray}
\label{disip_full:eq}
\hat{\mathcal{D}}&=\sum_{l,m}^{\{\rm{p,q}\}}\sum_{j,k=1}^n\bigg( ~M^{m,l}_{k,j}\hat{\mathcal{L}}^{l,m}_{j,k}+M^{m,l*}_{k,j}\hat{\mathcal{R}}^{l,m}_{j,k}\bigg).
\end{eqnarray}
By $(\bullet)^*$ we denote the complex-conjugation.
Plugging the relations (\ref{eq:fund_dissip}) in the equation (\ref{disip_full:eq}) we obtain an apprehensible, quadratic form of the Redfield dissipator
\begin{eqnarray}
\label{eq:dissipator_simpl}
\hat{\mathcal{D}}&=2(-\underline{\hat{c}}_0'\cdot{\bf M}^{\rm q,p}_{\rm i}\underline{\hat{c}}_0-2\underline{\hat{c}}_0'\cdot{\bf M}^{\rm q,q}_{\rm i}\underline{\hat{c}}_1+\underline{\hat{c}}_1'\cdot{\bf M}^{\rm p,p}_{\rm i}\underline{\hat{c}}_0+\underline{\hat{c}}_1'\cdot{\bf M}^{\rm p,q}_{\rm i}\underline{\hat{c}}_1)\\ \nonumber
&~~\,+\underline{\hat{c}}_0'\cdot{\bf M}^{\rm q,q}\underline{\hat{c}}_0'-\underline{\hat{c}}_0'\cdot({\bf M}^{\rm q,p})^*\underline{\hat{c}}_1'-\underline{\hat{c}}_1'\cdot{\bf M}^{\rm p,q}\underline{\hat{c}}_0'+\underline{\hat{c}}_1'\cdot({\bf M}^{\rm p,p})^*\underline{\hat{c}}_1'.
\end{eqnarray}
We shall use the following notation ${\bf M}^{\rm a,b}_{\rm r}={\rm Re}({\bf M}^{\rm a,b})$ and ${\bf M}^{\rm a,b}_{\rm i}={\rm Im}({\bf M}^{\rm a,b})$ for real and imaginary part of the matrices ${\bf M}^{\rm a,b}$, ${\rm a,b}\in \{{\rm p,q}\}$. Inserting now the forms (\ref{eq:liouv_unit}) and (\ref{eq:dissipator_simpl}) in the definition (\ref{eq:Liouville}) and defining the super-operator vector  $\underline{\hat{b}}=(\underline{\hat{c}}_0,\underline{\hat{c}}_1,\underline{\hat{c}}_0',\underline{\hat{c}}_1')^{\rm T}$ we arrive at the compact form of the complete Liouvillean
\begin{eqnarray}
\label{eq:lyap}
&\hat{\mathcal{L}}=\hat{\underline{b}}\cdot {\bf S} \hat{\underline{b}}-S_0\hat{\mathds{1}},\\ \nonumber
&{\bf S}=\left(
\begin{array}{cc}
{\bf 0}&{\bf -X} \\
{\bf -X}^{\rm T}&{\bf Y}\\
\end{array}
\right) ,\\ \nonumber 
&{\bf X}^{\rm T}=\left(
\begin{array}{cc}
{\bf R}+{\bf M}^{\rm q,p}_{\rm i},&{\bf Q}+{\bf M}^{\rm q,q}_{\rm i}\\
-{\bf P}-{\bf M}^{\rm p,p}_{\rm i},&-{\bf R}-{\bf M}^{\rm p,q}_{\rm i}\\
\end{array}
\right) ,\\ \nonumber 
&{\bf Y}=\frac{1}{2}\left(
\begin{array}{cc}
2{\rm i}{\bf Q}+{\bf M}^{\rm q,q}+({\bf M}^{\rm q,q})^{\rm T}, & -({\bf M}^{\rm q,p})^*-({\bf M}^{\rm p,q})^{\rm T}\\
-({\bf M}^{\rm q,p})^\dag-({\bf M}^{\rm p,q}), &-2{\rm i}{\bf P}+ ({\bf M}^{\rm p,p})^*+({\bf M}^{\rm p,p})^{\dag}\\
\end{array}
\right)={\bf Y}^{\rm T} ,\\ \nonumber 
&S_0=2{\rm Im}({\rm  Tr}\left( {\bf M}^{\rm p,q}-{\bf M}^{\rm q,p} \right)).
\end{eqnarray}
We name the part of the matrix ${{\bf X}}$ arising from the dissipator the coupling matrix and the matrix ${\bf Y}$ the driving matrix. The left stationary state of the Liouville equation (\ref{eq:master}) is simply the identity operator, $(1|\hat{\cal L} = 0$,  whereas the right stationary state determines the density matrix
$\rho_{\rm NESS}$ of the non-equilibrium stationary state, $\hat{\cal L}\rho_{\rm NESS}=0$. We now employ the main theorem of Ref. \cite{prosel10}. It states, that the right stationary state of a Liouvillean, which may be represented as a quadratic form in the creation ($\hat{c}_j'$) and annihilation ($\hat{c}_j$) super-operators, that satisfy canonical commutation relations, is determined by the covariance matrix (two point correlation function)
\begin{equation}
\label{eq:corr_def}
Z_{j,k}=(1|\hat{c}_j\hat{c}_k|{\rm NESS}\rangle.
\end{equation}
This correlation function is calculated from the continuous Lyapunov equation
\begin{equation}
\label{eq:lyap_def}
{\bf X}^{\rm T}\bf{Z}+{\bf ZX}={\bf Y}.
\end{equation}
The NESS is a Gaussian state, hence all expectation values can be calculated from the two point correlation function (\ref{eq:corr_def}) using the Wick theorem. In our case the correlation matrix ${\bf Z}$ gives us the momentum-coordinate correlations
\begin{eqnarray}
{\bf Z}=\left(\begin{array}{cc}
{\bf Z}^{\rm p,p},&({\bf Z}^{\rm q,p})^{\rm T}\\ 
{\bf Z}^{\rm q,p},&{\bf Z}^{\rm q,q}
\end{array}\right), \quad Z^{\rm a,b}_{j,k}={\rm tr}\{a_jb_k\rho_{\rm NESS}\},\quad {\rm a,b\in \{p,q\}}.
\end{eqnarray}
The correlation matrix ${\bf Z}$ is made real by adding the matrix 
${\bf Z}_0:=\frac{\rm i}{2}\sigma^{\rm x}\otimes\mathds{1}_n$. The map ${\bf Z}\rightarrow{\bf Z}+{\bf Z}_0$ implies ${\bf Y}\rightarrow{\bf Y}-{\bf X}^{\rm T}{\bf Z}_0-{\bf Z}_0{\bf X}$, therefore we have
\begin{eqnarray}
{\bf Z}=\left(\begin{array}{cc}
{\bf Z}^{\rm p,p},&(\tilde{\bf Z}^{\rm q,p})^{\rm T}\\ 
\tilde{\bf Z}^{\rm q,p},&{\bf Z}^{\rm q,q}
\end{array}\right),\quad \tilde{\bf Z}^{\rm q,p}={\bf Z}^{\rm q,p}+\frac{\rm i}{2}\mathds{1}_n,\\ \nonumber
{\bf Y}=\frac{1}{2}\left(
\begin{array}{cc}
{\bf M}_{\rm r}^{\rm q,q}+({\bf M}_{\rm r}^{\rm q,q})^{\rm T}, & -{\bf M}^{\rm q,p}_{\rm r}-({\bf M}^{\rm p,q}_{\rm r})^{\rm T}\\
-({\bf M}^{\rm q,p}_{\rm r})^{\rm T}-{\bf M}^{\rm p,q}_{\rm r}, &{\bf M}^{\rm p,p}_{\rm r}+({\bf M}^{\rm p,p}_{\rm r})^{\dag}\\
\end{array}
\right).
\end{eqnarray}
The coupling and the driving matrices are connected to the real and imaginary part of the bath matrices (\ref{eq:bath_matrix_def}), respectively. The Liouville super-operator in Redfield form does not necessary conserve positivity, therefore one should check if the obtained covariance matrix determines a bona fide density matrix. This is done by verifying the positivity of the covariance matrix $C_{j,k}={\rm tr}\{b_jb_k\rho_{\rm NESS}\}$, which is simply an unusual form of the Heisenberg uncertainty relation expressed for the covariance matrix \cite{simon94}. We introduced an operator vector $\underline{b}=(\underline{p},\underline{q})$.

\section{Harmonic oscillator chain}
\label{sec3.0}
In this section we use the derived theory to solve the example of harmonic oscillator chain coupled to heat baths at the ends of the chain.  The classical and quantum version of this problem have been studied in the literature \cite{dhar08,gaubut07} using the generalized quantum Langevin equation, derived in Ref. \cite{dhar} utilizing the Ford-Kac-Mazur formalism. A boundary condition dependent scaling of the heat flux for the gapless disordered chain has been observed. In Ref. \cite{gaubut07} the connection between heat current and entanglement has been analyzed. Here we propose a different approach to introduce the effect of the heat baths. We use the slightly modified Redfield dissipator, namely, we extend the lower integration limit in the equation dissipator (\ref{eq:dissip_def}) to $-\infty$, which is the same as neglecting the principal value of the integral. This further simplification is necessary for the Gibbs state to be the stationary state, when all baths have equal temperature. It is not clear under which conditions this approximation is valid, however it is necessary to obtain physically meaningful heat reservoirs (see Ref. \cite{njp2} for more discussion on this issue).

The Hamiltonian of a harmonic chain with nearest neighbor interaction may be expressed as a sum of local Hamiltonians and some boundary Hamiltonian 
\begin{equation}
H=\sum_{j=1}^{n-1}H_j+H_{\rm B}.
\end{equation}
The local Hamiltonians $H_j$ contain the contribution from the kinetic energy, on-site potential, and the interaction energy, whereas the boundary Hamiltonian consists of the kinetic and the potential part on the ends of the chain and the boundary condition term
\begin{eqnarray}
&H_j=\frac{1}{4}\left(p_j^{2}+p_{j+1}^{2}+\omega_j^2q_j^{2}+\omega_{j+1}^2q_{j+1}^{2}\right)+\frac{k_j}{2}\left(\frac{q_j}{\sqrt{m_j}}-\frac{q_{j+1}}{\sqrt{m_{j+1}}}\right)^{2},\\ \nonumber
&H_{\rm B}=\frac{1}{4}\left(p_1^{2}+p_{n}^{2}+\omega_1^2q_1^{2}+\omega_{n}^2q_{n}^{2}\right)+\frac{k'_1}{2m_1}q_1^{2}+\frac{k'_n}{2m_n}q_n^{2}.
\end{eqnarray} 
Employing a compact matrix notation we get
\begin{eqnarray}
&H=\frac{1}{2}\underline{p}\cdot\underline{p}+\frac{1}{2}\underline{q}\cdot{\bf Q}\underline{q},\\ \nonumber
&{\bf Q}=\left( \begin{array}{cccc}
\frac{k_1'+k_1}{m_1}+\omega_1^2&-\frac{k_1}{\sqrt{m_1m_2}},&\cdots,&0\\
-\frac{k_1}{\sqrt{m_1m_2}},&\frac{k_1+k_2}{m_2}+\omega_2^2,&\ddots,&\vdots\\ 
\vdots,&\ddots,&\ddots,&-\frac{k_{n-1}}{\sqrt{m_{n-1}m_n}}\\
0,&\cdots,&-\frac{k_{n-1}}{\sqrt{m_{n-1}m_n}},&\frac{k_{n-1}+k_n'}{m_n}+\omega_n^2\\
\end{array}\right).
\end{eqnarray}
We consider a natural choice of the coupling operators
\begin{eqnarray}
\!\!\!\!\!\!\!\!\!\!\!\!X_{\rm L}=\sqrt{\epsilon_{\rm L}}q_1,\quad X_{\rm R}=\sqrt{\epsilon_{\rm R}}q_n, \quad \epsilon_{\rm R}=\frac{\epsilon'_{\rm R}}{m_n},\quad \epsilon_{\rm R}=\frac{\epsilon'_{\rm R}}{m_n}
\end{eqnarray}
and two baths with possible different temperatures and general spectral functions
\begin{equation}
\Gamma^{\beta_j}_{j,k}(\omega)=\delta_{j,k}\frac{{\rm sign}({\omega})|\omega|^\nu}{\exp(\omega\beta_j)-1}.
\end{equation}
Three different types of the heat bath according to the parameter $\nu$ are possible: a) sub-ohmic bath ($\nu<1$), b) ohmic bath ($\nu=1$), and c) super-ohmic bath ($\nu>1$). We use the values $\nu=\frac{1}{2}$ and $\nu=2$ for the sub-ohmic and the super-ohmic case, respectively. The main goal of this section is to examine the behavior of heat current defined by the continuity equation in the bulk \footnote{The reasoning leading to equation (\ref{eq:heat_current}) for the heat current is the same as in \cite{njp2}.}
\begin{eqnarray}
\label{eq:heat_current}
\!\!\!\!\!\!\!\!\!\!\!\!\!\!\!\!\!\!\!\!\!\!\!\! 
&0=\frac{\rm d}{{\rm d}t}{\rm tr} (H_j \rho_{\rm NESS})
={\rm tr}\{{\rm i}[H,H_j]\rho_{\rm NESS}\}=J_{j-1,j}+J_{j+1,j},\\ \nonumber
\!\!\!\!\!\!\!\!\!\!\!\!\!\!\!\!\!\!\!\!\!\!\!\! 
&J_{j-1,j}={\rm tr}\left\{{\rm i}[H_{j-1},H_j]\rho_{\rm NESS}\right\}=-\frac{k_{j-1}}{m_{j}}{\rm tr}\left\{\left(p_{j}q_{j}-\frac{\rm i}{2}-p_{j}q_{j-1}\sqrt{\frac{m_j}{m_{j-1}}}\right)\rho_{\rm NESS}\right\},
\end{eqnarray}
where $J_{j-1,j}$ denotes the current from site $j-1$ to site $j$. From the antisymmetry constraint for the correlation matrix $\tilde{\bf Z}^{\rm qp}$, which we shall derive shortly, follow vanishing expectation values $\langle p_jq_j+q_jp_j\rangle=0$. Therefore the equation for the heat current simplifies to
\begin{eqnarray}
J=J_{j-1,j}=\frac{k_{j-1}}{\sqrt{m_{j}m_{j-1}}}{\rm tr}\left\{p_{j}q_{j-1}\rho_{\rm NESS}\right\}.
\end{eqnarray}

In the remaining part of this paper we investigate the problem of heat transport through the harmonic oscillator chain in the framework of third quantization for bosonic Redfield master equation. First, we recover some results for the homogeneous chains, then we focus on the scaling law of the heat current in the disordered chain. 

\subsection{Homogeneous chain}
\label{sec3.1}
In this section we study the homogeneous chain, where all coupling constants, masses and on-site potentials are equal; $k_j=k,~m_j=m,~\omega_j=\omega_0$ for all $j$ . Thus, only two parameters determine the behavior of the system, namely the on site potential $\omega_0^2$ and the coupling frequency $\omega_{\rm c}=\sqrt{\frac{k}{m}}$. The Hamiltonian is
\begin{eqnarray}
&H=\sum_{j=1}^n\left(\frac{p_j^2}{2}+\frac{\omega_0^2q_j^2}{2}\right)+\sum_{j=1}^{n-1}\frac{k}{2m}(q_{j}-q_{j+1})^2+\frac{k}{2m}(q_1^2+q_n^2), \\ \nonumber
&H=\frac{1}{2}\underline{p}\cdot\underline{p}+\frac{1}{2}\underline{q}\cdot{\bf Q}\underline{q},\\ \nonumber
&{\bf Q}=\omega^2\left( \begin{array}{cccc}
2&-1,&\cdots,&0\\
-1,&2,&\ddots,&\vdots\\ 
\vdots,&\ddots,&\ddots,&-1\\
0,&\cdots,&-1,&2\\
\end{array}\right)+\omega_0^2\mathds{I}_n.
\end{eqnarray}
In this section we restrict ourselves to fixed boundary conditions $k'_1=k'_n=k=1$, whereas in later discussion of the disordered chain we refer also to free boundary conditions, $k'_1=k'_n=0$.
The coordinate part of the Hamiltonian is diagonalizable by sine-Fourier transformation
\begin{eqnarray}
&{\bf Q}={\bf U} \Omega {\bf U}^\dag,\\ \nonumber
&U_{i,j}=\sqrt{\frac{2}{n+1}}\sin\left(\frac{ij\pi}{n+1}\right),\quad\Omega_{i,j}=\left(\omega_0^2+2\omega^2-2\omega^2\cos\left(\frac{i\pi}{n+1}\right)\right)\delta_{i,j}.
\end{eqnarray}
Normal coordinate and momentum operators are obtained by the transformations $\underline{q}'=\underline{q}\cdot{\bf U}$, $\underline{p}'=\underline{p}\cdot{\bf U}$. The Hamiltonian is then diagonalized via the usual transformation to creation and annihilation operators $q'_j=(a_j^\dag+a_j)/\sqrt{2\lambda_j}, ~\,p'_j={\rm i}\sqrt{\lambda_j}(a_j^\dag-a_j)/\sqrt{2}.$ We can make this transformations for a general harmonic chain where the columns of the matrix ${\bf U}$ are the right eigenvectors and $\lambda_j$ are the corresponding eigenvalues of the matrix ${\bf Q}$.  For the homogeneous chain we have ${\bf U}^\dag={\bf U}^{\rm T}={\bf U}$ and $\lambda_j=\Omega_{j,j}$. The Heisenberg propagator in the normal basis contains a coordinate and a momentum part 
\begin{eqnarray}
& q_j'(t)=f^{\rm q'}_j(t)q_j'+f^{\rm p'}_j(t)p_j'\\ \nonumber
& f^{\rm q'}_j=\frac{1}{2}\left({\rm e}^{{\rm i}\omega_j t}+{\rm e}^{-{\rm i}\omega_j t} \right) , \qquad f^{\rm p'}_j=\frac{\rm i}{\omega_j}\left({\rm e}^{-{\rm i}\omega_j t}-{\rm e}^{{\rm i}\omega_j t} \right).
\end{eqnarray}
The time dependent coupling operators in the normal basis are
\begin{eqnarray}
x_j^{\rm L}&=\sqrt{\frac{2\epsilon_{\rm L}}{n+1}}\sin\left(\frac{j\pi}{n+1}\right),\quad
X_{\rm L}(t)&=\underline{x}^{\rm L}\cdot ({\rm diag}(\underline{f}^{\rm q'}(t))\underline{q}'+{\rm diag}(\underline{f}^{\rm p'}(t))\underline{p}'),\\ \nonumber
x_j^{\rm R}&=\sqrt{\frac{2\epsilon_{\rm R}}{n+1}}\sin\left(\frac{j n \pi}{n+1}\right),\quad
X_{\rm R}(t)&=\underline{x}^{\rm R}\cdot ({\rm diag}(\underline{f}^{\rm q'}(t))\underline{q}'+{\rm diag}(\underline{f}^{\rm p'}(t))\underline{p}').
\end{eqnarray}
The problem now is to solve the continuous Lyapunov equation
\begin{eqnarray}
\label{eq:lyapun_chain}
&{\bf X}^{\rm T}{\bf Z}+{\bf ZX}={\bf Y},\\ \nonumber
&{\bf X}^{\rm T}=\left(\begin{array}{cc}
{\bf M}^{\rm q'p'}_{\rm i}, & \frac{{\Omega}}{2}\\
-\frac{\mathds{1}_n}{2},&{\bf 0}
\end{array}\right),\quad
{\bf Y}=\frac{1}{2}\left(\begin{array}{cc}
{\bf M}^{\rm q'q'}_{\rm r}+({\bf M}^{\rm q'q'}_{\rm r})^{\rm T}, & {\bf 0}\\
{\bf 0},&{\bf 0}
\end{array}\right).
\end{eqnarray}
Bath matrices ${\bf M}^{\rm q'q'}, {\bf M}^{\rm q'p'}$ are obtained from the equation (\ref{eq:bath_matrix_def}) applying the Fourier transformation $\Gamma^\beta(\omega)=\frac{1} {2\pi}\int_{-\infty}^{\infty}{\rm d}t\exp(-{\rm i}\omega t)\Gamma^\beta(t)$ and the Kubo-Martin-Schwinger (KMS) condition $\Gamma^{\beta}(-\omega)=\exp(\beta\omega)\Gamma^{\beta}(\omega)$
\small
\begin{eqnarray}
&{\bf M}^{\rm q'q'}=\frac{\pi}{2}\underline{x}^{\rm L}\otimes\underline{x}^{\rm L}{\rm diag}\left(\left(1+{\rm e}^{\beta_{\rm L}\lambda_j} \right)\Gamma^{\beta_{\rm L}}(\lambda_j)\right)+\frac{\pi}{2}\underline{x}^{\rm R}\otimes\underline{x}^{\rm R}{\rm diag}\left(\left(1+{\rm e}^{\beta_{\rm R}\lambda_j} \right)\Gamma^{\beta_{\rm R}}(\lambda_j)\right),\\ \nonumber
&{\bf M}^{\rm q'p'}=\frac{{\rm i}\pi}{2}\underline{x}^{\rm L}\otimes\underline{x}^{\rm L}{\rm diag}\left(\left( \frac{{\rm e}^{\beta_{\rm L}\lambda_j}-1}{\lambda_j}\right)\Gamma^{\beta_{\rm L}}(\lambda_j) \right)+\frac{{\rm i}\pi}{2}\underline{x}^{\rm R}\otimes\underline{x}^{\rm R}{\rm diag}\left(\left( \frac{{\rm e}^{\beta_{\rm R}\lambda_j}-1}{\lambda_j}\right)\Gamma^{\beta_{\rm R}}(\lambda_j) \right).
\end{eqnarray}
\normalsize
In the case of equal temperatures of the baths the Gibbs state should be a stationary state of the modified Redfield master equation. Therefore, it is a good exercise and consistency check to calculate the equilibrium solution (for  $\beta_{\rm L}=\beta_{\rm R}=\beta$) of the Lyapunov equation (\ref{eq:lyapun_chain})
\small
\begin{eqnarray}
Z^{\rm p'p'}_{i,j}=\delta_{i,j}\frac{\lambda_j\left(1+{\rm e}^{\beta\lambda_j} \right)}{2\left({\rm e}^{\beta\lambda_j} -1\right)},\quad Z^{\rm q'q'}_{i,j}=\delta_{i,j}\frac{\left(1+{\rm e}^{\beta\lambda_j} \right)}{2\lambda_j\left({\rm e}^{\beta\lambda_j} -1\right)},\quad {\bf Z}^{\rm q'p'}=-\frac{\rm i}{2}\mathds{1}_n,\quad{\bf Z}^{\rm p'q'}=\frac{\rm i}{2}\mathds{1}_n.
\label{eq:gibbs_ness}
\end{eqnarray}
\normalsize
Transforming the last result to the annihilation-creation operator correlation matrix we indeed obtain a well known formula for the occupation number of a harmonic oscillator in the equilibrium (thermal) state at inverse temperature $\beta$
\begin{eqnarray}
Z^{\rm a^\dag a}_{j,k}&=\frac{1}{2}\left( \lambda_j Z^{\rm q'q'}_{j,k}+\frac{Z^{\rm p'p'}_{j,k}}{\lambda_j}+{\rm i}Z^{\rm p'q'}_{j,k}-{\rm i}Z^{\rm q'p'}_{j,k}\right)\\ \nonumber
&=\frac{\delta_{j,k}}{2}\left(\frac{\left(1+{\rm e}^{\beta\lambda_j} \right)}{\left({\rm e}^{\beta\lambda_j} -1\right)}-1\right)=\frac{\delta_{j,k}}{{\rm e}^{\beta\lambda_j} -1}.
\end{eqnarray}
This solution is independent of the coupling operators $X_{\rm L}, X_{\rm R}$ and detailed structure of the bath correlation functions, however, they should satisfy the KMS condition. Therefore, the same correlation matrix is obtained for a general harmonic oscillator chain with heat baths of equal temperatures. We prove that by expanding the Lyapunov equation (\ref{eq:lyapun_chain})
\begin{eqnarray}
\label{eq:lyap2}
&{\bf M}^{\rm q'q'}+({\bf M}^{\rm q'q'})^{\rm T}={\bf M}_{\rm i}^{\rm q'p'}{\bf Z}^{\rm p'p'}+\frac{1}{2}{\Omega}(\tilde{\bf Z}^{\rm q'p'})^{T}+{\bf Z}^{\rm p'p'}({\bf M}^{\rm q'p'}_{\rm i})^{T}+\frac{1}{2}\tilde{\bf Z}^{\rm q'p'}{\Omega},\\ \nonumber
&{\bf 0}={\bf M}^{\rm q'p'}_{\rm i}\tilde{\bf Z}^{\rm q'p'}+\frac{1}{2}{\Omega}{\bf Z}^{\rm q'q'}-\frac{1}{2}{\bf Z}^{\rm p'p'},\\ \nonumber
&{\bf 0}=\tilde{\bf Z}^{\rm q'p'}+(\tilde{\bf Z}^{\rm q'p'})^{\rm T}.
\end{eqnarray}
These equations must be solved consistently with the symmetry conditions $({\bf Z}^{\rm q'q'})^{\rm T}={\bf Z}^{\rm q'q'}$ and $ ({\bf Z}^{\rm p'p'})^{\rm T}={\bf Z}^{\rm p'p'}$. From the last equation (\ref{eq:lyap2}) follows the antisymmetry of the matrix $\tilde{\bf Z}^{\rm q'p'}={\bf Z}^{\rm q'p'}+\frac{\rm i}{2}\mathds{1}$. Further, by simplifying the bath matrices for the case of equal temperatures
\begin{eqnarray}
\!\!\!\!\!\!\!\!\!\!\!\!\!\!\!{\bf M}^{\rm q'q'}&=\frac{\pi}{2}\left(\underline{x}^{\rm L}\otimes\underline{x}^{\rm L}+\underline{x}^{\rm R}\otimes\underline{x}^{\rm R}\right){\rm diag}\left(\left(1+{\rm e}^{\beta\lambda_j} \right)\Gamma^{\beta}(\lambda_j)\right),\\ \nonumber
\!\!\!\!\!\!\!\!\!\!\!\!\!\!\!{\bf M}^{\rm q'p'}&=\frac{{\rm i}\pi}{2}\left(\underline{x}^{\rm L}\otimes\underline{x}^{\rm L}+\underline{x}^{\rm R}\otimes\underline{x}^{\rm R}\right){\rm diag}\left(\left( \frac{{\rm e}^{\beta_{\rm R}\lambda_j}-1}{\lambda_j}\right)\Gamma^{\beta_{\rm R}}(\lambda_j) \right)
\end{eqnarray}
it is straightforward to see that correlations (\ref{eq:gibbs_ness}) solve the equation (\ref{eq:lyap2}).

\begin{figure}[h!!!]
\includegraphics[scale=0.67]{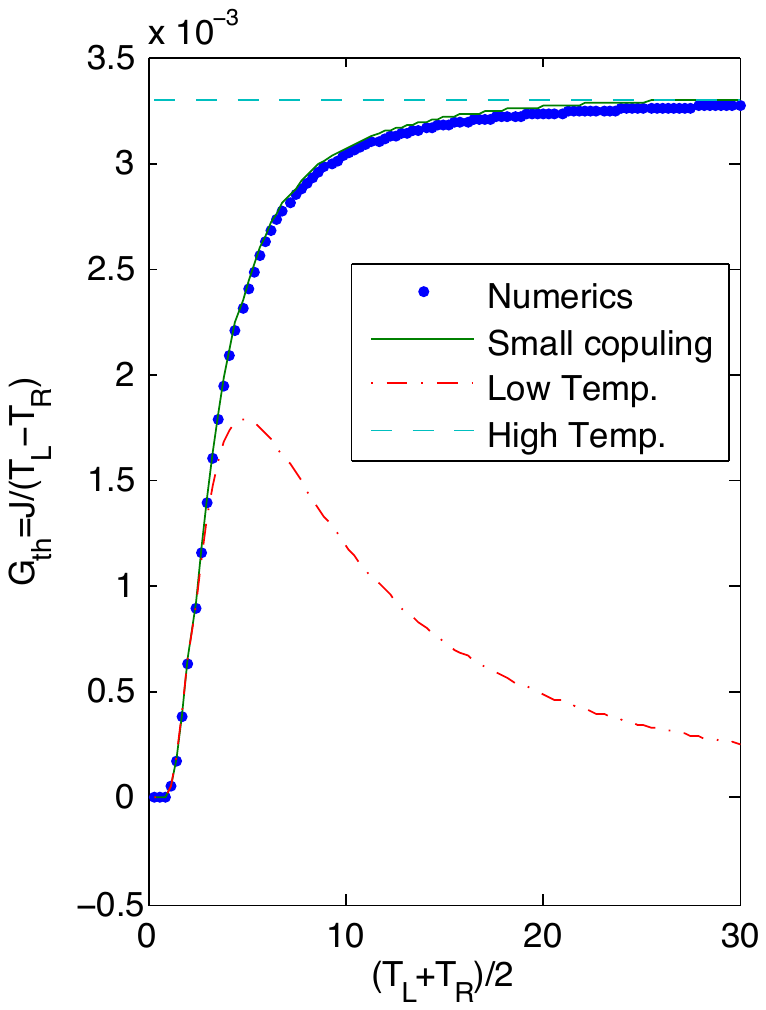}
\includegraphics[scale=0.67]{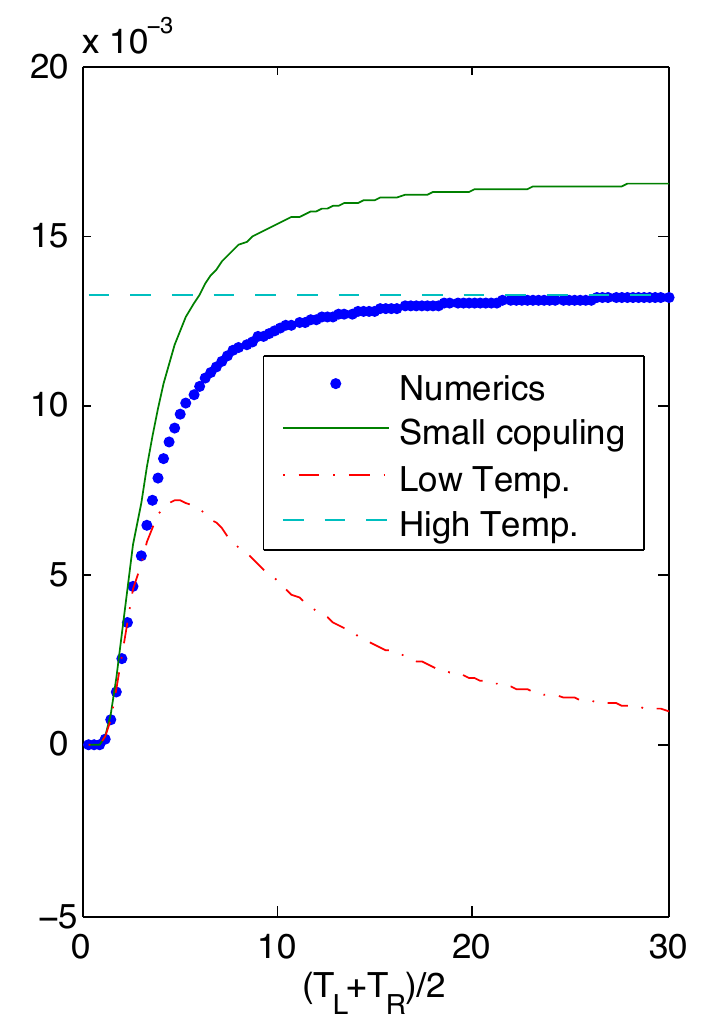}
\includegraphics[scale=0.67]{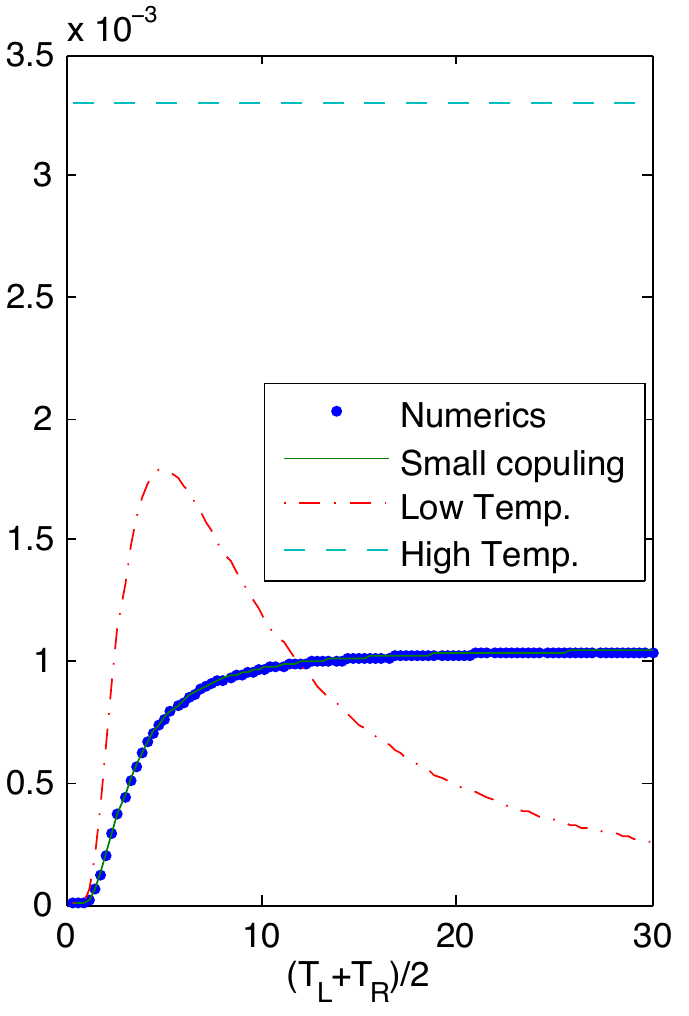}
\caption{Comparison of numerically calculated thermal conductance and analytic solutions in the high temperature (\ref{eq:prevodnost_high}), and the low temperature (\ref{eq:prevodnost_low}) limits and the small coupling limit (\ref{eq:prevodnost_small_eps}) for a homogeneous chain. We remark that analytic solutions for high and low temperature limits are asymptotically correct for all coupling strengths, which can be seen from the middle graph ($\epsilon'=5\times10^{-2}$ and ohmic bath), whereas the small coupling solution is valid for all correlation functions, as shown on the right graph ($\epsilon'=10^{-2}$, sub-ohmic bath). The left graph is plotted for $\epsilon'=10^{-2}$ and ohmic bath. Other model parameters are: $n=20,~ (T_{\rm L} - T_{\rm R})/(T_{\rm L}+T_{\rm R})=0.01,~ \omega_0=10,~ \omega_{\rm c}=2/3,~ \epsilon_{\rm L}'=\epsilon_{\rm R}'=\epsilon$.}
\label{fig:prevodnost}
\end{figure}

\subsection{Homogeneous chain - general solution}
\label{sec3.2}
In this section we provide a general solution for the homogeneous oscillator chain with ohmic bath correlation function $\Gamma^{\beta}(\omega)=\frac{\omega}{\exp(\beta\omega)-1}$.  In the case of fixed boundary conditions and equal, symmetric coupling $\epsilon_{\rm L}=\epsilon_{\rm R}=\epsilon$ it is possible to find an exact explicit expression for the correlation matrix. Straightforward but tedious calculation, in which we exploit the particular form of the bath matrices, shows that the solution of equations (\ref{eq:lyap2}), rewritten in real space coordinates, can be sought in the form of the ansatz
\small
\begin{eqnarray}
\label{eq:corr_sol_ohmic}
&\tilde{Z}^{\rm qp}_{i,j}={\rm sign}(i-j)z_{|i-j|},\quad z_n=z_0=0,\\ \nonumber
&Z^{\rm pp}_{i,j}= [{\bf U}{\rm diag}(\underline{\tilde{\Gamma}}^+){\bf U}]_{i,j}+\epsilon \left\{ 
  \begin{array}{cc}
    z_{|i+j-2|}-az_{|i+j-1|}+z_{|i+j|}, & \quad \mbox{if $i+j<n+1$}\\
    -z_{|2n-i-j|}+az_{|2n-i-j-1|}-z_{|2n-i-j-2|}, & \quad \mbox{if $i+j>n+1$}\\
0,& \quad \mbox{otherwise}\\
  \end{array} \right.
\end{eqnarray}
\normalsize
We defined the bath functions
\begin{equation}
\label{eq:ansatz}
\tilde{\Gamma}_j^\pm=\frac{1}{2}\left(\frac{\lambda_j\left(1+{\rm e}^{\beta_{\rm L}\lambda_j} \right)}{\exp(\beta_{\rm L}\lambda_j)-1}\pm\frac{\lambda_j\left(1+{\rm e}^{\beta_{\rm R}\lambda_j} \right)}{\exp(\beta_{\rm R}\lambda_j)-1}\right)
\end{equation}
and the constant $a=2+\omega_0^2/\omega_{\rm c}^2.$  The first part of the correlation matrix ${\bf Z}^{\rm pp}$ in equation (\ref{eq:ansatz}) has the same form as in the equilibrium case, except the temperatures can be different. On  the other hand, the second part is calculated from the equations (\ref{eq:lyap2}) and the ansatz for $\tilde{\bf Z}^{qp}$. The Lyapunov equation is now simplified to a linear system of $n-1$ equations for the unknown coefficients $z_1,\ldots z_{n-1}.$
For a large number of oscillators $n\rightarrow\infty$ the covariance matrix is connected to a solution of the continuous fraction equation
\begin{eqnarray}
x=\frac{\tilde{a}+\sqrt{\tilde{a}^2-4}}{2},\quad \tilde{a}=\frac{\omega^2_{\rm c}}{\epsilon^2}+a.
\end{eqnarray}
The heat current $J=\omega^2_{\rm c}z_1$ is then given by
\begin{equation}
z_1=\frac{1}{\epsilon^2x^2}(2 x \phi_1+\phi_2+\phi_3+\ldots \phi_{n-1})=\frac{2(x \phi_1-\phi_1)}{\epsilon^2x^2}+\frac{1}{\epsilon^2x^2}\sum_{k=1}^{n-1}\phi_k,
\end{equation}
where the values $\phi_1,\phi_2,\ldots \phi_n$ are calculated as follows
\begin{eqnarray}
\phi_k=\sum_{j=1}^n\frac{2\epsilon\tilde{\Gamma}^-_j}{n+1}\sin\left(\frac{\pi j}{n+1}\right)\sin\left(\frac{k\pi j}{n+1}\right).
\end{eqnarray}
In the high-temperature limit the expression for the heat current is simplified to
\begin{equation}
J_{\rm high}\approx\frac{\omega_{\rm c}^2}{2x\epsilon}(T_{\rm L}-T_{\rm R}).
\label{eq:prevodnost_low}
\end{equation}
This is an expected behavior since for high temperatures the thermal conductance $G_{\rm th}=J/(T_{\rm L}-T_{\rm R})$ should be independent of the temperature, which follows from classical mechanical considerations. In the low temperature limit we expect that thermal conductance will go to zero. We can simplify the thermal conductance for large on-site potential ($\omega_0\gg\omega_{\rm c}$) and small relative temperature difference $|T_{\rm R}-T_{\rm L}|/(T_{\rm L}+T_{\rm R}) \ll 1$, to
\begin{eqnarray}
G_{\rm th}\approx\frac{2\omega^2_{\rm c}\omega^2_0}{\epsilon x}\frac{{\rm e}^{-2\omega_0/(T_{\rm L}+T_{\rm R})}}{(T_{\rm L}+T_{\rm R})^2}.
\label{eq:prevodnost_high}
\end{eqnarray}

\subsection{Recursive solution}
For a general spectral function of the form $\Gamma^{\beta}(\omega)={\rm sign}(\omega)\frac{|\omega|^{\nu}}{\exp(\beta\omega)-1}$ and equal couplings it is possible to find a recursive solution for the correlation matrix in terms of a perturbative ansatz
\begin{equation}
{\bf Z}^{\rm p'p'}=\sum_{j=0}^\infty\epsilon^{2j}{\bf Z}^{{\rm p'p'}(2j)},\quad {\bf Z}^{\rm q'q'}=\sum_{j=0}^\infty\epsilon^{2j}{\bf Z}^{{\rm q'q'}(2j)},\quad{\bf Z}^{\rm p'p'}=\sum_{j=0}^\infty\epsilon^{2j+1}{\bf Z}^{{\rm p'p'}(2j+1)}.
\end{equation}
The form of the ansatz can be guessed if we rewrite the equations $(\ref{eq:lyap2})$ order by order. Starting from the lowest order we have:
\begin{eqnarray}
&Z^{\rm p'p'(0)}_{j,k}=\frac{\delta_{j,k}}{4|\lambda|^{\nu-1}}\left(\left(1+{\rm e}^{\beta_{\rm L}\lambda_j} \right)\Gamma^{\beta_{\rm L}}(\lambda_j)+\left(1+{\rm e}^{\beta_{\rm R}\lambda_j} \right)\Gamma^{\beta_{\rm R}}(\lambda_j)\right)=\frac{\tilde{\Gamma}^+_j}{2},\\ \nonumber
&Z^{\rm q'q'(0)}_{j,k}=\frac{\delta_{j,k}}{4|\lambda|^{\nu+1}}\left(\left(1+{\rm e}^{\beta_{\rm L}\lambda_j} \right)\Gamma^{\beta_{\rm L}}(\lambda_j)+\left(1+{\rm e}^{\beta_{\rm R}\lambda_j} \right)\Gamma^{\beta_{\rm R}}(\lambda_j)\right)=\frac{\tilde{\Gamma}^+_j}{2\lambda_j^2},\\ \nonumber
&\tilde{Z}^{\rm q'p'(1)}_{j,k}=(1-(-1)^{k+j})\frac{(M^{\rm qq}_{j,k}+M^{\rm qq}_{k,j})}{\lambda^2_j-\lambda^2_k}.
\end{eqnarray}
Each subsequent order is then calculated from the previous one using the recursion relations
\begin{eqnarray}
\label{eq:recursive_solution}
&Z^{\rm q'q'(n)}_{j,k}=\frac{2}{\lambda_k^2-\lambda_j^2}\left\{[{\bf M}^{\rm q'p'}\tilde{\bf Z}^{{\rm q'p'}(n-1)}]_{j,k}-[{\bf M}^{\rm q'p'}\tilde{\bf Z}^{{\rm q'p'}(n-1)}]_{k,j}\right\}, \\ \nonumber
&{\bf Z}^{\rm p'p'(n)}=2{\bf M}^{\rm q'p'}\tilde{\bf Z}^{{\rm q'p'}(n-1)}+{\Omega}{\bf Z}^{{\rm q'q'}(n-1)},\\ \nonumber
&\tilde{Z}_{j,k}^{\rm q'p' (n+1)}=\frac{1-(-1)^{j+k}}{\lambda_j^2-\lambda_k^2} \left\{[{\bf M}^{\rm q'p'}{\bf Z}^{{\rm p'p'}(n)}]_{j,k}+[{\bf M}^{\rm q'p'}{\bf Z}^{{\rm p'p'}(n)}]_{k,j}\right\}.
\end{eqnarray}
We can show that in each order for arbitrary initial condition $\tilde{\bf Z}^{{\rm qp}(n-1)}$ respecting the antisymmetry condition  one can calculate the next order of of the correlations ${\bf Z}^{{\rm pp}(n)}$ and ${\bf Z}^{{\rm qq}(n)}$, which also satisfy the symmetry conditions and from which the next correction for the coordinate-momentum part of the correlation matrix can be calculated uniquely. There might be problems on the diagonal due to the difference of the eigenvalues, but the special form of the bath matrix $M^{\rm qp}_{j,k}$, which is nonzero only when the sum $j+k$ is odd, ensures that all diagonal corrections apart from ${\bf Z}^{{\rm pp}(0)}$ and ${\bf Z}^{{\rm qq}(0)}$ vanish. We emphasize that  the solution (\ref{eq:recursive_solution}) is usefull only for $\epsilon_{\rm L}=\epsilon_{\rm R}=\epsilon<1$. 

In the first order in the coupling $\epsilon$ it is possible to find simple explicit results for the kinetic energy profile and the heat current. We connect the local kinetic energy to the local temperature at the same site 
\begin{equation}
T_j=Z^{\rm p,p}_{j,j},
\label{eq:temp}
\end{equation} 
which is in principle justified only if the system obeys the condition of local thermal equilibrium. This is definitely not the case for harmonic systems, so our ``temperature" field $T_j$ which is simply defined by (\ref{eq:temp}) should be understood only as a technical term.
We find constant temperature profile for long chains and large on site potential $\omega_0\gg\omega$
\begin{equation}
\!\!\!\!\!\!\!\!\!\!\!\!Z^{\rm p'p'}_{k,k}\approx\frac{\tilde{\Gamma}^+_{n/2}}{4}\approx\frac{1}{4}\left(\frac{\omega_0\left(1+{\rm e}^{\beta_{\rm L}\omega_0} \right)}{\exp(\beta_{\rm L}\omega_0)-1}+\frac{\omega_0\left(1+{\rm e}^{\beta_{\rm R}\omega_0} \right)}{\exp(\beta_{\rm R}\omega_0)-1}\right).
\end{equation}
We get the same form of the matrix $\tilde{\bf Z}^{\rm q'p'}$ as in the ohmic bath case, hence, the heat current in the weak coupling limit is
\begin{equation}
J=\epsilon\sum_{j=1}^n\sin^2\left(\frac{j\pi}{n+1}\right)|\lambda_j|^{\nu-1}\tilde{\Gamma}^-_j.
\end{equation}
In the large on-site potential limit ($\omega_0\gg\omega_{\rm c}$) the current is simplified to
\begin{equation}
J\approx\frac{\epsilon}{2}|\lambda_{n/2}|^{\nu-1}\tilde{\Gamma}^-_{n/2}\approx\frac{\epsilon\omega_0^\nu}{4}\left(\frac{1+{\rm e}^{\beta_{\rm L}\omega_0} }{\exp(\beta_{\rm L}\omega_0)-1}-\frac{1+{\rm e}^{\beta_{\rm R}\omega_0} }{\exp(\beta_{\rm R}\omega_0)-1}\right).
\label{eq:prevodnost_small_eps}
\end{equation}
Comparison of analytical solutions in the low-temperature (\ref{eq:prevodnost_low}) and high-temperature (\ref{eq:prevodnost_high}) limits and small-coupling limit (\ref{eq:prevodnost_small_eps}) with the numerical results for the thermal conductance is shown in Fig. (\ref{fig:prevodnost}).

\begin{figure}[!htb]

\vbox{
	\includegraphics[scale=0.66]{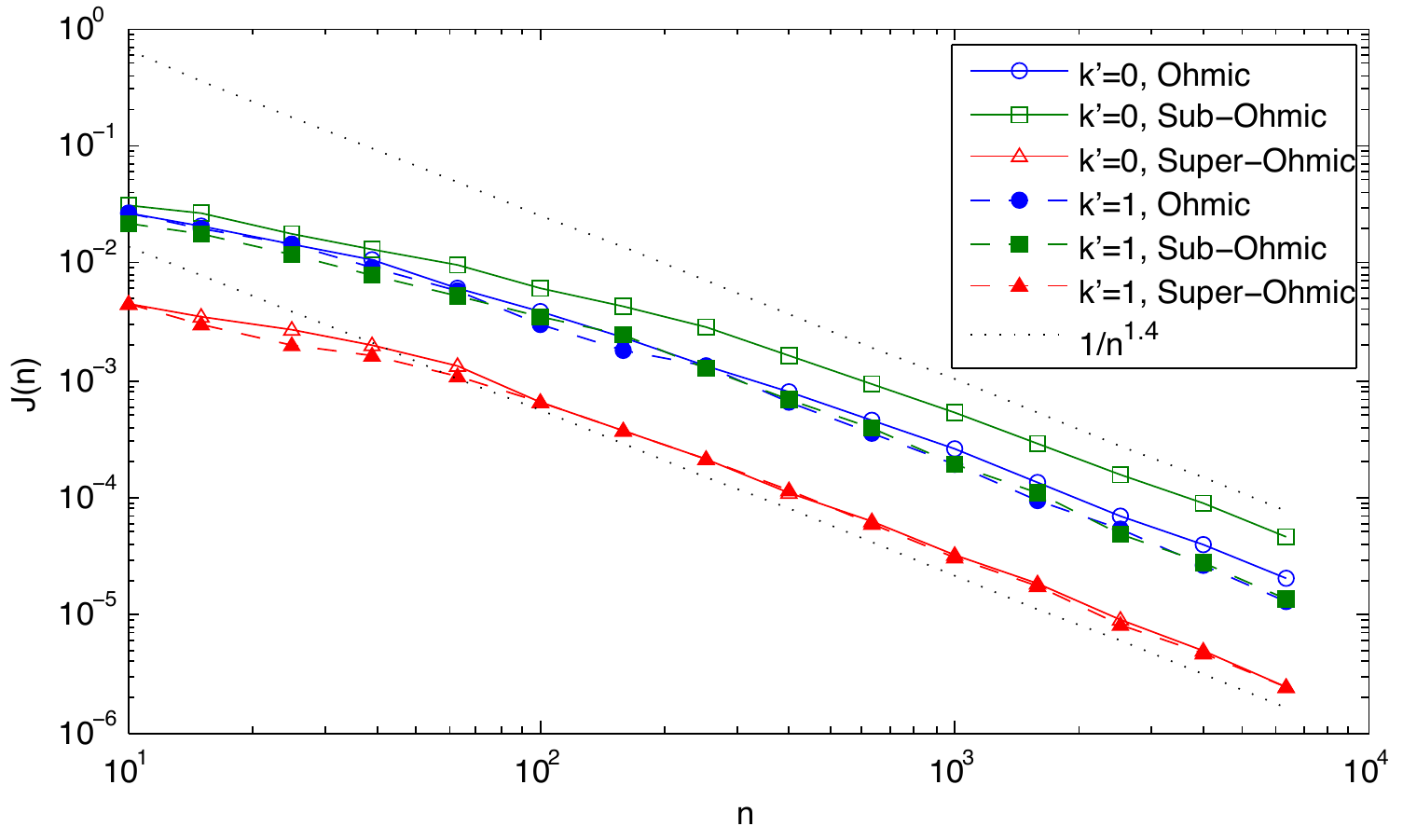}
	
	\includegraphics[scale=0.66]{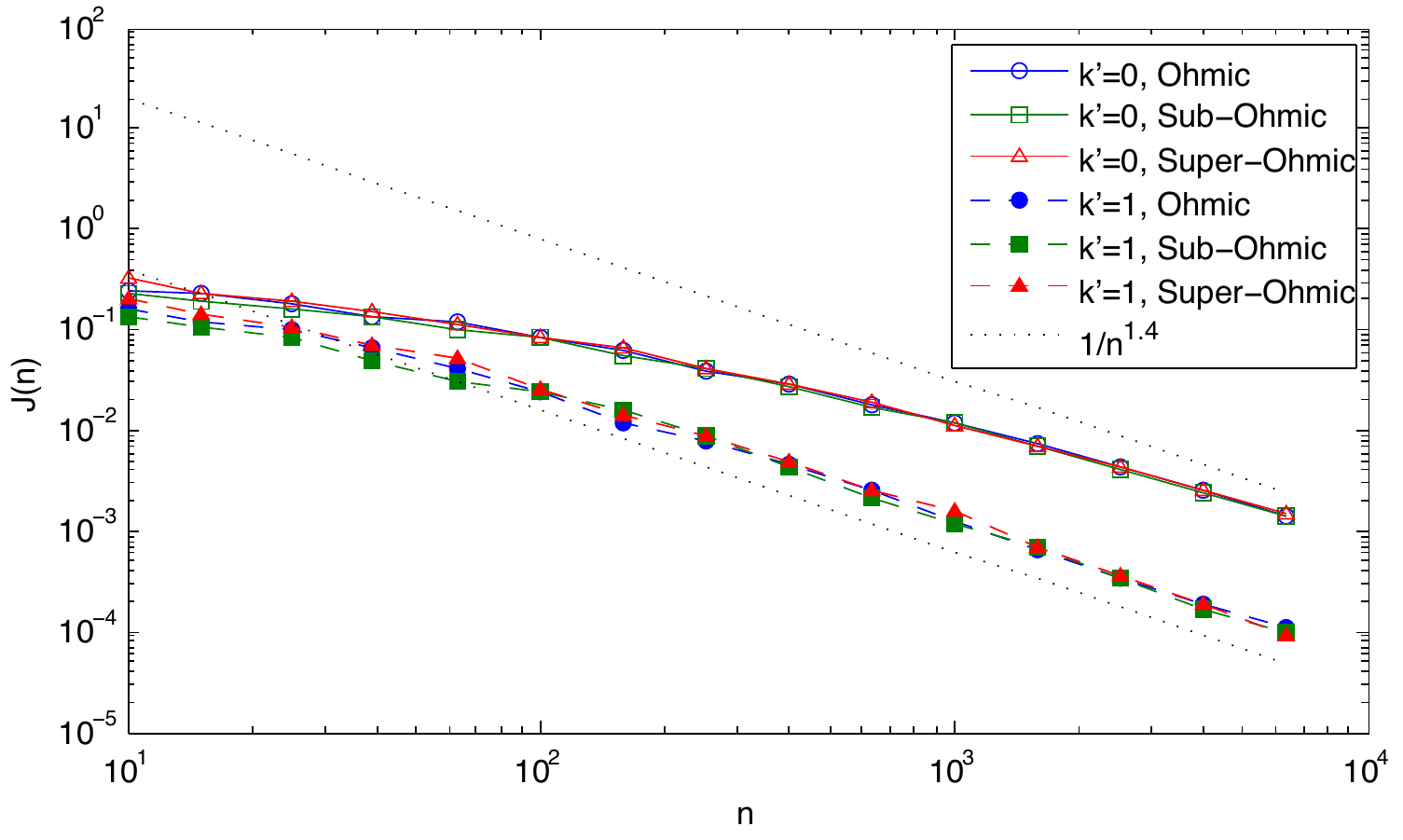}	
	
	\includegraphics[scale=0.66]{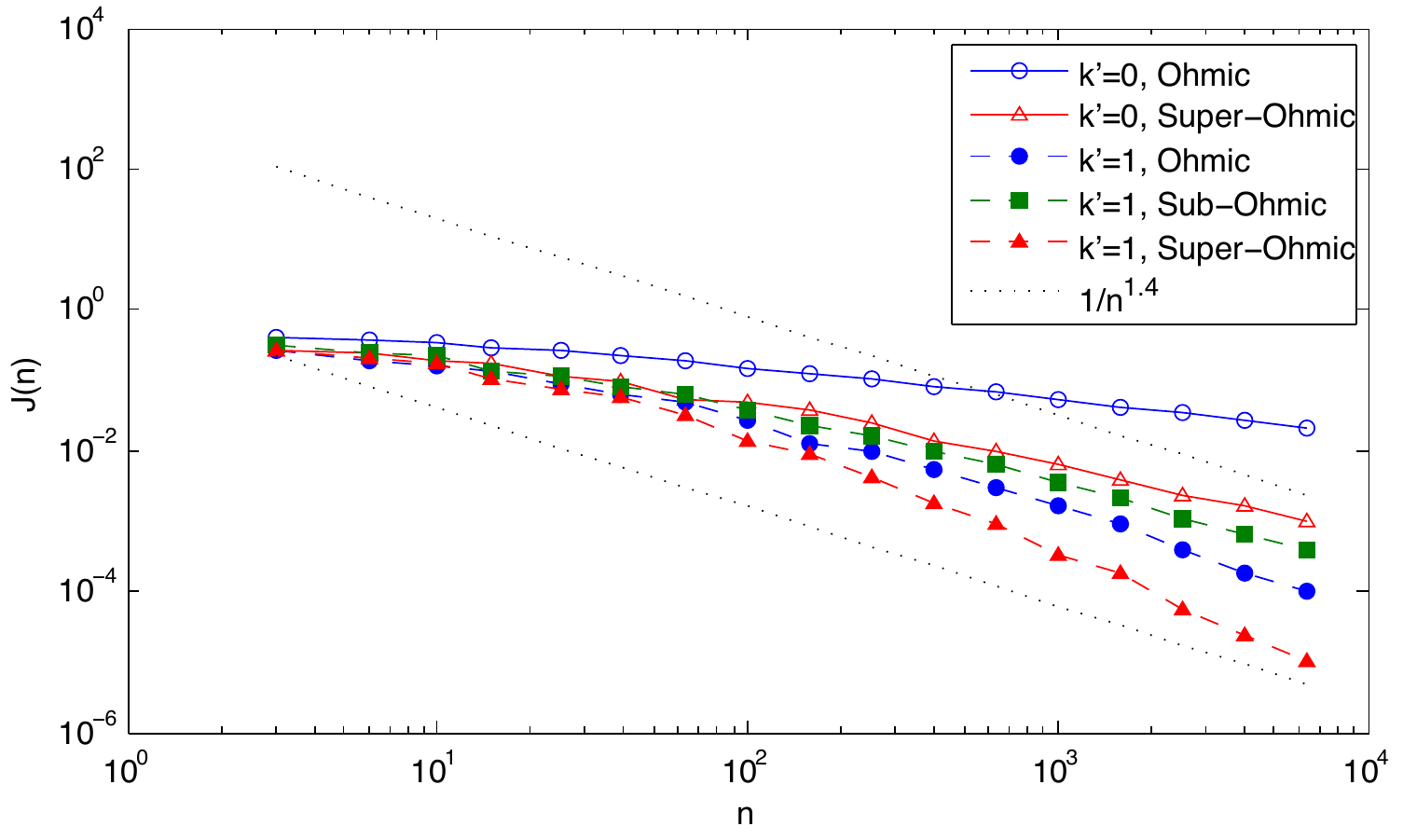}	
}	
	
	\caption{Averaged heat current scaling for different bath spectral functions (ohmic, Sub-ohmic, Super-ohmic) and boundary conditions ($k'=0,1$). The first panel (top) is for $\omega_0=10$, the middle panel for $\omega_0=1$, and the bottom panel for $\omega_0=10^{-6}$ (almost gapless). The thermodynamic behavior  for nonzero $\omega$ does not depend on the properties of the baths and varying model parameters ($\omega_0, k'$). The estimated asymptotic behavior is $\sim n^{-1.4}$. The statistical error is approximately the size of the symbols. Model parameters: $\delta=0.8,~\omega=2/3,~\epsilon_{\rm L}'=0.3,~\epsilon_{\rm R}'=0.1,~\beta_{\rm L}=0.1,~\beta_{\rm R}=1.0.$}
	\label{fig:current_scal_dis}
\end{figure}

\begin{figure}[!htb]
	\includegraphics[scale=0.75]{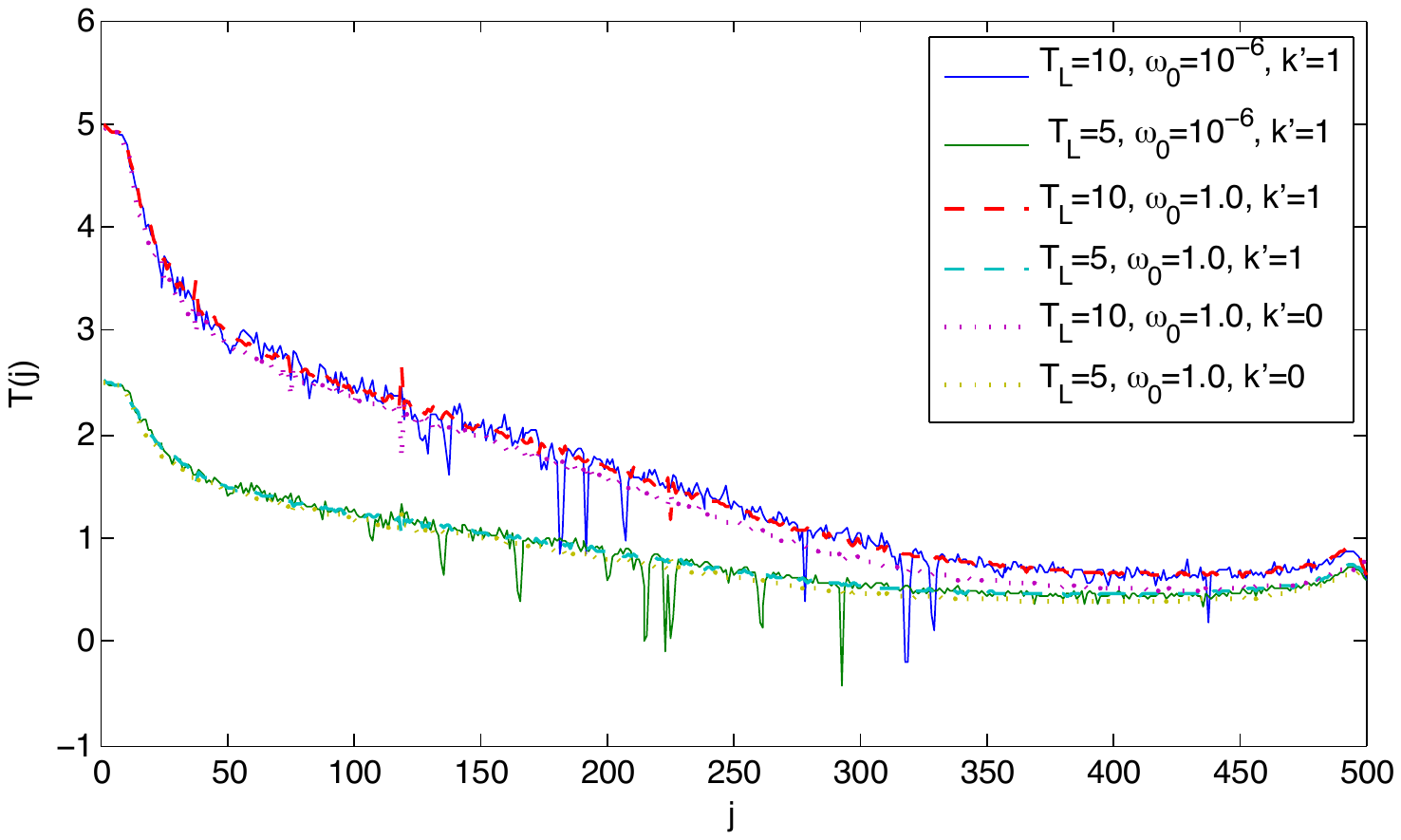}
	\caption{Averaged temperature profile (\ref{eq:temp}) for different temperatures and on site potential and boundary conditions. Average the is made over 1000 realizations. 
	Model parameters: $\delta=0.3,\omega=2/3,~\epsilon_{\rm L}'=0.3,~\epsilon_{\rm R}'=0.1,~\beta_{\rm R}=1.0$ and ohmic bath spectral function. 
	}
	\label{fig:t_profile}
\end{figure}

\subsection{Disordered chain}
\label{subs:dis_chain}

The homogeneous harmonic chains exhibit ballistic transport. Either one of the two different effects is needed for a sub-ballistic transport, say of a finite (non-zero) temperature gradient and finite (non-infinite) heat conductivity in the thermodynamic limit, namely: the interaction, which causes the dissipation of the phonons, or the disorder, which localizes the phonons. As interaction represents a whole new spectrum of problems and phenomena (see e.g. \cite{prozni10}), we shall focus in this subsection to disorder in harmonic chains. However disorder often leads even to a sub-diffusive (insulating) behavior in the thermodynamic limit. 
Disorder can be introduced by (i) randomly changing the masses, (ii) the on-site frequencies, or (iii) the coupling constants. We shall confine ourselves to the first type (i), sometimes referred to as isotopic disorder. The masses are chosen randomly  $m_j=m_0(1+\delta \xi_j)$, where $m_0$ denotes the mean mass, $\delta\in[0,1]$ denotes the relative width of the deviation, and $\xi_j$ is a random number uniformly distributed in the  interval [-1,1]. In the literature both, the isotopic disorder \cite{dhar08} and the disorder in the coupling constants \cite{gaubut07}, have been studied in the context of quantum Langevin equation. A finite temperature gradient and thermal conductivity in the thermodynamic limit have been observed \cite{gaubut07}.

Here, by employing our formalism, we show some new numerical results which differ from previous numerical conjectures \cite{dhar, dhar08,gaubut07}. Investigating the disordered harmonic oscillator Hamiltonian with free boundary conditions by using the quantum Langevin equation has been found that the thermodynamic behavior of the system depends on the properties of the heat baths \cite{dhar}. 
However, using our Redfield operator space formalism we find heat current scaling independent of the bath spectral functions and coupling parameters. Numerical calculation of the heat current for different bath correlation functions, coupling strengths and boundary conditions is presented in Fig. (\ref{fig:current_scal_dis}). Note that in order to obtain reliable data the heat current and the temperature profile are calculated as an average over many realizations of disorder. For the gapped cases, $\omega_0 \neq 0$, we find asymptotically, as $n\to \infty$, anomalous sub-diffusive power law scaling
\begin{equation}
J(n) \propto n^{-\alpha}, \quad {\rm where} \quad \alpha = 1.4.
\end{equation}
On the other hand, data seem less conclusive for the gapless case (in order to avoid numerical instabilities with our method we have chosen $\omega_0 = 10^{-6}$) where the scaling $J(n)$ is either sensitive to the bath correlation function or the accessible system sizes $n$ are still too small, and the behaviors $J(n)$ are not yet asymptotic.
In Fig.~ (\ref{fig:t_profile}) we plot the temperature profiles for the ohmic baths and different on-site petentials and boundary conditions. It is perhaps remarkable to observe non-monotonicities in the profiles which are possible due to absence of local thermal equilibrium.

\section{Summary}
We generalized the method of the bosonic quantization in the Fock space of operators (third quantization) to a general markovian master equation. Lyapunov equation for the coordinate-momentum correlation function (the covariance matrix), which fully determines the non-equilibrium stationary state, has been obtained using the concept of left and right multiplication maps in the operator space. 

The formalism that we had derived was used to investigate the heat transport in the one-dimensional quantum harmonic oscillator chain with modified Redfield dissipators. An additional approximation turned out to be necessary in order to obtain consistent heat baths, which yield -- in the case of equal temperatures of the reservoirs -- the correct Gibbs state as the steady state solution of the master equation. For the homogeneous harmonic chain two analytical results for the covariance matrix were derived. The first one,  given by equations (\ref{eq:corr_sol_ohmic}), is valid for the ohmic bath and equal, possibly large, strengths of the coupling. The second is given in a perturbative, recursive form (\ref{eq:recursive_solution}) and is valid for general spectral function and again equal couplings. This solution involves complicated expressions for higher orders, therefore it is useful only in the small coupling limit, where already the first correction is adequate. Simplified expressions for the heat conductance and the temperature profile were derived. As expected from the classical results and previous quantum calculations \cite{dhar08} for the ordered chain we obtained vanishing temperature gradient and constant heat current in the thermodynamic limit. In the more interesting disordered case we find heat current scaling independent of the coupling, the bath properties, the on-site potential (as long as the phonon spectrum is gapped), and of the boundary conditions. The asymptotical decrease is faster than $1/n$ (Fourier law), in fact it seems asymptotically $\propto n^{-1.4}$, which corresponds to an insulator.
\section*{Acknowledgements}
We acknowledge useful discussions with Daniel Kosov and financial support by the Programme P1-0044, and the Grant J1-2208, of the Slovenian Research Agency (ARRS). 

\bibliography{redfield_oscillator_aipproc}

\begin{thebibliography}{22}
\expandafter\ifx\csname natexlab\endcsname\relax\def\natexlab#1{#1}\fi
\providecommand{\enquote}[1]{``#1''}
\expandafter\ifx\csname url\endcsname\relax
  \def\url#1{\texttt{#1}}\fi
\expandafter\ifx\csname urlprefix\endcsname\relax\def\urlprefix{URL }\fi
\providecommand{\eprint}[2][]{\url{#2}}

\bibitem[Casati(1985)]{giulio}
G.~Casati, \emph{Foundations of Physics} \textbf{16}, 51 (1985).

\bibitem[Ford(1992)]{ford}
J.~Ford, \emph{Physics Reports} \textbf{213}, 217 (1992).

\bibitem[Bonetto et~al.(2000)]{bonetto}
F.~Bonetto, J.~L. Lebowitz, and L.~Rey-Bellet, \emph{"Fourier law: A challenge
  to theorists" in Mathematical Physics 2000}, 128-150, Imperial College Press,
  2000.

\bibitem[Livi et~al.(2003)]{lepri}
R.~Livi, A.~Politi, and S.~Lepri, \emph{Phisics Reports} \textbf{377}, 1
  (2003).

\bibitem[Prosen and Campbell(2005)]{david}
T.~Prosen, and D.~K. Campbell, \emph{Chaos} \textbf{15}, 015117 (2005).

\bibitem[Dhar(2008)]{dhar08}
A.~Dhar, \emph{Advances in Physics} \textbf{57}, 457 (2008).

\bibitem[Haug and Jauho(1998)]{keldysh}
H.~Haug, and A.~P. Jauho, \emph{Quantum Kinetics in Transport and Quantum
  Kinetics in Transport and Optics of Semiconductors}, Springer-Verlag, Berlin,
  1998.

\bibitem[Datta(1995)]{landauer}
S.~Datta, \emph{Electronic transport in mesoscopic systems}, Cambridge
  University Press, Cambridge, 1995.

\bibitem[Ford et~al.(1988)]{qle}
G.~W. Ford, J.~T. Lewis, and R.~F. O'Connell, \emph{Phys. Rev. A} \textbf{37},
  4419 (1988).

\bibitem[Prosen(2008)]{njp}
T.~Prosen, \emph{New J. Physics} \textbf{10}, 043026 (2008).

\bibitem[Prosen and {\v Z}unkovi{\v c}(2010)]{njp2}
T.~Prosen, and B.~{\v Z}unkovi{\v c}, \emph{New J. Physics} \textbf{12}, 025016
  (2010).

\bibitem[Prosen(2010)]{pro10}
T.~Prosen, \emph{J. Stat. Mech.} \textbf{2010}, P07020 (2010).

\bibitem[Prosen and Seligman(2010)]{prosel10}
T.~Prosen, and T.~H. Seligman, \emph{J. Phys. A: Math. Theor.} \textbf{43},
  392004 (2010).

\bibitem[Reider et~al.(1967)]{lieb}
Z.~Reider, J.~L. Lebowitz, and E.~Lieb, \emph{J. Math. Phys.} \textbf{8}, 1073
  (1967).

\bibitem[Dhar(2001)]{dharprl}
A.~Dhar, \emph{Phys. Rev. Lett.} \textbf{86}, 5882 (2001).

\bibitem[Dhar and Shastry(2003)]{dhar}
A.~Dhar, and B.~S. Shastry, \emph{Phys. Rev. B} \textbf{67}, 195405 (2003).

\bibitem[Gaul and Buttner(2007)]{gaubut07}
C.~Gaul, and H.~Buttner, \emph{Phys. Rev. E} \textbf{76}, 011111 (2007).

\bibitem[Breuer and Petruccione(2002)]{breuer}
H.~P. Breuer, and F.~Petruccione, \emph{The theory of open quantum systems},
  Oxford University Press, Cambridge, 2002.

\bibitem[Saito(2003)]{saito}
K.~Saito, \emph{Europhys Lett.} \textbf{62}, 34 (2003).

\bibitem[Saito et~al.(2000)]{satami00}
K.~Saito, S.~Takesue, and S.~Miyashita, \emph{Phys. Rev. E} \textbf{61}, 3
  (2000).

\bibitem[Simon et~al.(1994)]{simon94}
R.~Simon, N.~Mukunda, and B.~Dutta, \emph{Phys. Rev. A} \textbf{49}, 1567
  (1994).

\bibitem[Prosen and {\v Z}nidari{\v c}(2010)]{prozni10}
T.~Prosen, and M.~{\v Z}nidari{\v c}, \emph{Phys. Rev. Lett.} \textbf{105},
  060603 (2010).

\end{thebibliography}

 \end{document}